\documentclass[10pt,twocolumn,superscriptaddress,showpacs,aps,amsmath,amssymb,nofootinbib,pra]{revtex4-1}
\pdfoutput=1

\usepackage{bm}
\usepackage{textcomp}
\usepackage{bbm}

\usepackage{tikz}
\usetikzlibrary{calc}

\usepackage{hyperref}
\hypersetup{%
pdftitle={Emulating the one-dimensional Fermi-Hubbard model by a double chain of qubits},%
pdfauthor={Jan-Michael Reiner, Michael Marthaler, Jochen Braum\"uller, Martin Weides, and Gerd Sch\"on},%
colorlinks=true,%
linkcolor=blue,%
urlcolor=blue,%
citecolor=blue
}

\newcommand{\crop}{\vspace*{-2em}}
\newcommand{\e}{\mathrm{e}}

\begin{document}

\title{Emulating the one-dimensional Fermi-Hubbard model by a double chain of qubits}

\author{Jan-Michael Reiner}
\affiliation {Institut f\"ur Theoretische Festk\"orperphysik, Karlsruhe Institute of Technology (KIT), 76131 Karlsruhe, Germany}

\author{Michael Marthaler}
\affiliation {Institut f\"ur Theoretische Festk\"orperphysik, Karlsruhe Institute of Technology (KIT), 76131 Karlsruhe, Germany}
      
\author{Jochen Braum\"uller}
\affiliation {Physikalisches Institut, Karlsruhe Institute of Technology (KIT), 76131 Karlsruhe,
Germany}

\author{Martin Weides}
\affiliation {Physikalisches Institut, Karlsruhe Institute of Technology (KIT), 76131 Karlsruhe,
Germany}
\affiliation {Physikalisches Institut, Johannes Gutenberg University Mainz, 55128 Mainz,
Germany}

\author{Gerd Sch\"on}
\affiliation {Institut f\"ur Theoretische Festk\"orperphysik, Karlsruhe Institute of Technology (KIT), 76131 Karlsruhe, Germany}
\affiliation {Institute of Nanotechnology, Karlsruhe Institute of Technology (KIT),  76021 Karlsruhe, Germany}

\date{\today}

\begin{abstract}
The Jordan-Wigner transformation maps a one-dimensional (1D) spin-$1/2$ system onto a fermionic model without spin degree of freedom. A double chain of quantum bits with $XX$ and $ZZ$ couplings of neighboring qubits along and between the chains, respectively, can be mapped on a spin-full 1D Fermi-Hubbard model. The qubit system can thus be used to emulate the quantum properties of this model. We analyze  physical implementations of such analog quantum simulators, including one based on transmon qubits, where the $ZZ$ interaction arises due to an inductive coupling and the $XX$ interaction due to a capacitive interaction. We propose protocols to gain confidence in the results of the simulation through measurements of local operators.
\end{abstract}

\pacs{03.67.-a, 71.10.Fd, 85.25.-j}%

\maketitle

\section{Introduction}

Simulations of quantum systems on a classical computer are limited by the fact that the Hilbert space grows exponentially with the number of particles considered. The problem could be solved if a universal, \emph{digital} quantum computer of sufficient size would be available. Short of this solution one can try to solve the problem by using \emph{analog} quantum simulators (AQS), or \emph{emulators}. For this purpose one needs to design artificial systems governed by Hamiltonians which can be mapped onto those of the quantum systems to be studied. An example with more than 300 qubits with sufficient coherence properties has been realized recently in experiments~\cite{350_Spin_Simulator}. In spite of such progress, the strategy of emulating further model Hamiltonians by artificially created systems remains a challenging task. 

Of particular interest, because of the difficulties they pose, are {\sl fermionic} systems. Great progress in the simulation of such systems has been made by using cold atom gases~\cite{Guan_RevModPhys13} or trapped ions~\cite{Trapped_Fermi_2002,Trapped_Fermi_2006,Frustrated_Spins}. The Fermi-Hubbard model has been studied with atoms in optical traps \cite{Buluta02102009}. In these experiments one can use actual fermionic particles, which allows a straightforward mapping of the artificial system onto the fermionic model of interest. On the other hand, in these systems the available range of coupling strengths, or equivalently of effective temperatures, is limited. In addition, the individual control and readout of atoms still poses problems, which makes it difficult to read out correlation functions.

Controlled access to individual ``particles'' is routinely achieved for systems consisting of superconducting qubits. In one dimension, the properties of (spin-less) Fer\-mions can be mapped  by the Jor\-dan\--Wigner transformation onto those of a chain of qubits. In extension to this, the on-site interaction and the hopping of a spin-full 1D Fermi-Hubbard model can be modeled by using a double chain (ladder) structure of qubits with $XX$ (more precisely spin-flip-type) and $ZZ$ couplings of neighboring qubits along the chains and between the chains, respectively \cite{Shastry1986}. This mapping has been exploited in numerical treatments of the problem \cite{Znidaric2012}. Here we discuss how these systems can be used for the purpose of quantum emulations and analyze different physical realizations of the qubits, including charge qubits and transmons. Josephson junction arrays of charge qubits \cite{jctarray1,jctarray2,chargequbit1,chargequbit2,chargequbit3} are conceptually the simplest model and allow all needed operations, but they are known to suffer from the random offset charge problem. Transmon qubits \cite{Koch_Transmon} are more stable against noise and, therefore, are frequently favored. Their energy splitting can be tuned via a  SQUID loop providing the effective Josephson coupling; the $ZZ$ interaction arises due to a mutual inductance, and the $XX$ interaction via a capacitive interaction. On the other hand, when using transmons we find restrictions on the accessible range of parameters of the Fermi-Hubbard model. For both systems it is routinely possible to measure local operators and correlators. By performing several of such measurements one can also gain confidence in the quality of the simulation.

The paper is organized as follows: In the next section we review how the Hamiltonian of the spin-full one-dimensional (1D) Fermi-Hubbard model can be mapped onto the double chain of qubits. Then we discuss the physical realizations of the qubit double chain by superconducting qubits. Finally we suggest methods of read-out and initialization as well as measurements which allow testing the quality of the emulation. Physical properties of transmon qubits and their coupling required for the simulation are discussed in detail in several appendices.

\section{Mapping fermions onto Qubits}

A single fermionic state, which can be occupied or non-occupied, can be mapped onto the two basis states of a spin-1/2 particle or qubit. However, for a set of spins the raising and lowering operators $\sigma^\pm_j$ do not obey the anticommutation relations of the corresponding fermionic creation and annihilation operators $c^\dagger_j$ and $c^{\phantom \dagger}_j$. Instead, for $j \neq j'$ one finds commutation relations $[\sigma^-_j, \sigma^+_{j'}]=[\sigma^+_j, \sigma^+_{j'}]=[\sigma^-_j, \sigma^-_{j'}]=0$, since the excitations of qubits are bosonic.

For 1D systems a correct mapping between fermions and qubits is provided by the Jordan-Wigner transformation. Consider an ordered set of qubits and define $c^{\phantom \dagger}_{j} = \mathrm{e}^{\mathrm i \pi \lambda_j} \sigma^-_j$ with $\lambda_j = \sum_{k=1}^{j-1} \sigma^+_k \sigma^-_k$. The operators $c_j$ defined in this way are fermionic, satisfying the anticommutation relations $\{ c^{\dagger}_{j}, c^{\phantom \dagger}_{j'} \} = \delta_{jj'}$ and $\{ c^{\phantom \dagger}_{j}, c^{\phantom \dagger}_{j'} \} = \{ c^\dagger_j, c^\dagger_{j'} \} = 0$. The exchange interactions between the fermions map onto the coupling of qubits via
\begin{align} \label{eq:Multi-Qubit-Interactions}
c^\dagger_j c^{\phantom \dagger}_{j'} = \sigma^+_j \mathrm{e}^{\mathrm i \pi (\lambda_j - \lambda_{j'})} \sigma^-_{j'} \, .
\end{align}
This relation shows that the coupling not only depends on the states of qubits $j$ and $j'$, but also on $\lambda_j - \lambda_{j'}$, i.e., on the states of all qubits between $j$ and $j'$. As a result, fermionic systems with two particle interactions in general map onto qubit systems with potentially complicated multi-qubit interactions. On the other hand, the Jordan-Wigner transformation also yields
\begin{equation}\label{eq:Jordan-Wigner-Identities}
\begin{gathered}
c_j^\dagger c^{\phantom \dagger}_{j} = \sigma_j^+ \sigma_j^- = \frac{1}{2} ( \sigma_j^z + 1 ), \\
c_j^\dagger c^{\phantom \dagger}_{j \pm 1} = \sigma_j^+ \sigma_{j \pm 1}^-.
\end{gathered}
\end{equation}
This allows for an easy mapping of 1D {\it spinless} fer\-mi\-onic systems onto a chain of spins as long as the exchange interaction couples only nearest neighbors~\cite{Anderson_Kondo_1D_Lattice,Nori_Majorana_Modes_in_Qubits}. 

\begin{figure}
\includegraphics[width=\columnwidth]{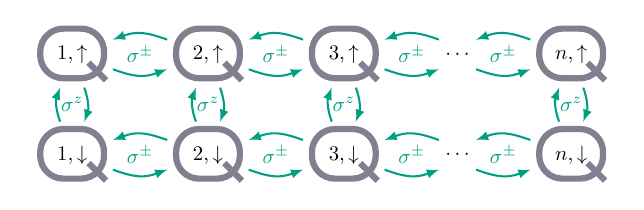}
\crop
\caption{
Layout of a qubit system consisting of two chains (labeled $\uparrow,\downarrow$) of $n$ qubits, where the neighboring qubits of each chain are coupled via a SWAP interaction of the form $\sigma^+_{j,s} \sigma^-_{j+1,s} + \mathrm{H.c.}$ The two chains from a ladder, with qubits belonging to the same rung coupled through terms involving $\sigma^z_{j,\uparrow} \sigma^z_{j,\downarrow}$. With the resulting Hamiltonian $H_\mathrm{QS}$~\eqref{eq:Hamiltonian-Emulator}  the qubit system  can serve as an analog quantum simulator (AQS) of the 1D Fermi-Hubbard model $H_\mathrm{FH}$~\eqref{eq:Hamiltonian-Fermi-Hubbard-Model}.
}
\label{fig:Qubit-System}
\end{figure}

Emulating $n$ interacting fermions {\it with spin-$\mathit{1/2}$} is possible by using the ladder-type array of qubits displayed in Fig.~\ref{fig:Qubit-System}~\cite{Shastry1986,Znidaric2012}. The qubit system consists of two chains of qubits of length $n$, with exchange interactions expressed by $\sigma^\pm$ operators between neighboring qubits in each chain. In addition, the two qubits of the different chains belonging to the same rung of the ladder are coupled by  $\sigma^z$ operators. Assuming that the qubit level spacings $\epsilon$ are all the same, and similarly the coupling strengths $g^x$ of the $\sigma^\pm$-type couplings as well as the strengths $g^z$ of the $\sigma^z$-type couplings, we arrive at the Hamiltonian 
\begin{align}\label{eq:Hamiltonian-Emulator}
H_\mathrm{QS} &= \sum_{j=1}^n \sum_{s=\uparrow,\downarrow} \frac{1}{2} \epsilon \sigma^z_{j,s} + g^z \sum_{j=1}^n \sigma^z_{j,\uparrow} \sigma^z_{j,\downarrow} \nonumber \\
&\quad + g^x \sum_{j=1}^{n-1} \sum_{s=\uparrow,\downarrow} (\sigma^+_{j,s} \sigma^-_{j+1,s} + \sigma^+_{j+1,s} \sigma^-_{j,s}).
\end{align}
The index $(j,s) \in \{1,\dots,n\}\times \{\uparrow,\downarrow\}$ refers to the qubit at the $j^\mathrm{th}$ position in the upper or lower chain. In order to perform the Jordan-Wigner transformation we need a consecutive, strictly one-dimensional numbering. It can be generated by the bijection $(j,\downarrow) \mapsto j$ and $(j,\uparrow) \mapsto j+n$. Using this and Eqs.~\eqref{eq:Jordan-Wigner-Identities} we can transform the three contributions in the Hamiltonian~\eqref{eq:Hamiltonian-Emulator}. The first sum is easy to treat since
\begin{equation}
\begin{gathered}
\sigma^z_{j,\downarrow} \mapsto \sigma^z_j = 2 c^\dagger_j c^{\phantom \dagger}_j - 1,\\
\sigma^z_{j,\uparrow} \mapsto \sigma^z_{j+n} = 2 c^\dagger_{j+n} c^{\phantom \dagger}_{j+n} - 1.
\end{gathered}
\end{equation}
The resulting $ZZ$ interaction between nonconsecutive qubits is unproblematic regarding the multi-qubit interactions of Eq.~\eqref{eq:Multi-Qubit-Interactions} since
\begin{align}
\sigma^z_{j,\downarrow} \sigma^z_{j,\uparrow} \mapsto \sigma^z_j \sigma^z_{j+n} = 4 (c^\dagger_j c^{\phantom \dagger}_j - \frac{1}{2}) (c^\dagger_{j+n} c^{\phantom \dagger}_{j+n} - \frac{1}{2}).
\end{align}
The exchange interaction, on the other hand, only occurs between nearest neighbors in our chosen order, and we find
\begin{equation}
\begin{gathered}
\sigma^+_{j,\downarrow} \sigma^-_{j \pm 1,\downarrow} \mapsto \sigma^+_{j} \sigma^-_{j \pm 1} = c^\dagger_j c^{\phantom \dagger}_{j \pm 1},\\
\sigma^+_{j,\uparrow} \sigma^-_{j \pm 1,\uparrow} \mapsto \sigma^+_{j+n} \sigma^-_{j \pm 1+n} = c^\dagger_{j+n} c^{\phantom \dagger}_{j \pm 1+n}.
\end{gathered}
\end{equation}
Using these relations in the Hamiltonian~\eqref{eq:Hamiltonian-Emulator}, and transforming the indices back through $j \mapsto (j,\downarrow)$ and $j+n \mapsto (j,\uparrow)$, one finds that the qubit system is equivalent to the Fermi-Hubbard model in one dimension with the Hamiltonian
\begin{align}\label{eq:Hamiltonian-Fermi-Hubbard-Model}
H_\mathrm{FH} &=  -\mu \sum_{j=1}^{n} \sum_{s = \uparrow,\downarrow} c^\dagger_{j,s} c^{\phantom \dagger}_{j,s} + U \sum_{j=1}^n c^\dagger_{j,\uparrow} c^{\phantom \dagger}_{j,\uparrow} c^\dagger_{j,\downarrow} c^{\phantom \dagger}_{j,\downarrow} \nonumber \\
&\quad - t \sum_{j=1}^{n-1} \sum_{s = \uparrow,\downarrow} (c^\dagger_{j,s} c^{\phantom \dagger}_{j+1,s} + c^\dagger_{j+1,s} c^{\phantom \dagger}_{j,s}).
\end{align}
The chemical potential $\mu$, the on-site energy $U$ and the transfer energy $t$ are related to the parameters of the qubit system via
\begin{align} \label{eq:Fermi-Hubbard-Parameters}
\mu = -\epsilon + 2 g^z, \quad U = 4 g^z, \quad \text{and } t = -g^x.
\end{align}

At this stage we conclude that the considered qubit system with the described $XX$ and $ZZ$  nearest-neighbor interactions is equivalent to the 1D Fermi-Hubbard mo\-del~\eqref{eq:Hamiltonian-Fermi-Hubbard-Model} {\sl including spin} and should allow emulating the latter. We draw attention to the fact that spin flip processes are not allowed. In the following sections we will investigate specific physical realizations of the qubits, including the questions what are the accessible ranges of parameters and what are the available tools for manipulation and measurement.

\section{Physical realizations of the qubit chains}

\begin{figure}
\includegraphics[width=.66\columnwidth]{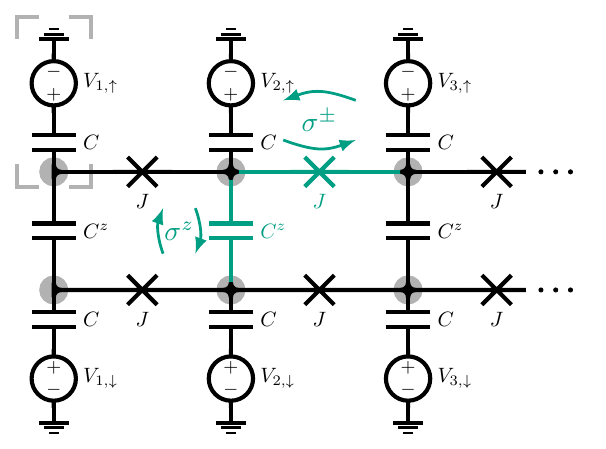}
\caption{
Circuit diagram for a physical realization of the quantum simulator of Fig.~\ref{fig:Qubit-System} by a ladder array of charge qubits (gray box). Superconducting islands (gray circles) within the same chain are coupled via Josephson junctions $J$, neighboring islands belonging to the two different chains are coupled via capacitances $C^z$. Each islands is connected via a capacitance $C$ to a control voltage $V_{j,s}$. Two charge states of the islands form the basis of the qubit. The circuit provides the desired couplings of Fig.~\ref{fig:Qubit-System} and allows tuning the chemical potential without further restriction, thus allowing any occupation of the fermionic states.
}
\label{fig:Qubit-Circuit-Josephson-Array}
\end{figure}

\begin{figure}
\includegraphics[width=\columnwidth]{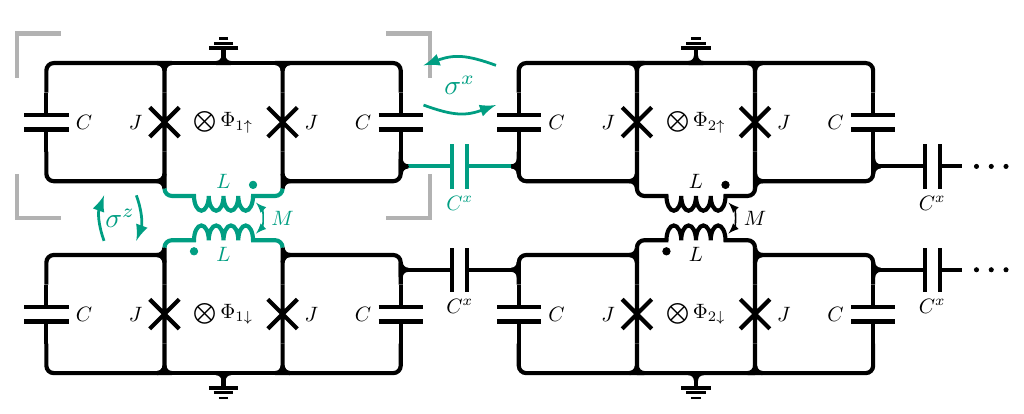}
\crop
\caption{
Superconducting circuit based on tunable transmon qubits (gray box). Each transmon consists of two Josephson junctions $J$ forming a loop with geometric inductance $L$; the junctions are shunted by capacitances $C$. Each individual transmon is tunable via applied magnetic fluxes $\Phi_{j,s}$ threading the loop. Additional  capacitances $C^x$ create an $XX$ coupling between two qubits. Coupling the inductances of two qubits via the mutual inductance $M$ produces a $ZZ$ interaction. The circuit is therefore a  realization of the qubit system of Fig.~\ref{fig:Qubit-System}.
}
\label{fig:Qubit-Circuit}
\end{figure}

The spin system shown in Fig.~\ref{fig:Qubit-System} with Hamiltonian $H_\mathrm{QS}$ given by Eq.~\eqref{eq:Hamiltonian-Emulator}, which can serve as an emulator of the 1D spin-full Fermi-Hubbard model, can be realized by superconducting circuits. One possibility is a realization based on an array formed by Josephson charge qubits \cite{jctarray1,jctarray2} as pictured in Fig.~\ref{fig:Qubit-Circuit-Josephson-Array} . Two nearly degenerate charge states of the superconducting islands are the basis of the charge qubit \cite{chargequbit1,chargequbit2,chargequbit3}. The islands are coupled via Josephson junctions and capacitances to neighboring ones.  The capacitances $C^z$ between islands belonging to the same rung give rise to the $ZZ$ interactions, while the Josephson junctions provide the exchange interactions along the two chains. The junctions have to be designed such that their intrinsic capacitances, which would lead to unwanted $ZZ$ interaction along the chains, are much smaller than $C^z$. Each island is further connected via a capacitance $C$ to a gate voltage $V_{j,s}$, which allows adjusting the level spacing of each qubit to the same value $\epsilon$. According to the relation~\eqref{eq:Fermi-Hubbard-Parameters} this allows tuning the chemical potential $\mu$ of the simulated Fermi-Hubbard model~\eqref{eq:Hamiltonian-Fermi-Hubbard-Model}. Equilibrium properties can thus be measured for a wide range of the chemical potential, which also allows studying the important case of half-filling. The values of $U$ and $t$ are fixed during fabrication, but they can be chosen in a wide range. Furthermore, by replacing the Josephson junctions by tunable SQUIDs, one can tune the parameter $t$.

While the realization via ideal Josephson charge qubits would allow the emulation of the Fermi-Hubbard model in the interesting parameter regime, and conceptually is most easily understood, it suffers from a serious problem. The  charge qubits are very sensitive to uncontrolled offset charges and background charge fluctuations. As a consequence it is very difficult to tune the system to  a homogeneous chemical potential. This problem has been recognized in the field on superconducting qubits, and the strategy was developed to explore other designs. 

An alternative design makes use of qubits with junctions based on phase slip processes~\cite{Mooij06,Mooij05}. Such phase slip qubits have been fabricated and have been shown to behave in a quantum coherent way~\cite{Astafiev}. For arrays built of such devices the disorder effects should be much weaker~\cite{Mooij15}. However, at this time it is too early to judge the quality and potential of these setups. Perhaps the prospects of using them for emulations will encourage further engagement into this new technology.  

A successful qubit design, favored nowadays by many experimentalists, is the  transmon~\cite{Koch_Transmon}. It is optimized to be less sensitive to charge noise. We therefore proceed to analyze a circuit based on transmon qubits as shown in Fig.~\ref{fig:Qubit-Circuit}. In this setup the $XX$ exchange interaction is provided by a capacitative coupling between transmons, and the $ZZ$-type interaction by a coupling via mutual inductances between the transmons.

In order to explain these couplings we shortly review the properties of a transmon; a detailed derivation is given in the Appendices~\ref{Appx:Tunable-Transmon}, \ref{Appx:ZZ-Coupling}, and~\ref{Appx:XX-Coupling}. One single tunable transmon is built from two Josephson junctions with critical current $I_\mathrm{c}$ (here assumed to be equal) and phase differences $\phi_\mathrm{l}$ and $\phi_\mathrm{r}$ across them. Each junction is shunted by a capacitance $C$, and they form a loop with low geometric inductance $L$. We introduce the external phase $\phi_\mathrm{e} = \frac{2e}{\hbar} \Phi_\mathrm{e}$ associated with the external flux $\Phi_\mathrm{e}$ through the loop containing the junctions, and the energy scales $E_C = \frac{e^2}{2C}$, $E_J = \frac{\hbar}{2e} I_\mathrm{c}$, and $E_L = \frac{\hbar^2}{e^2 L}$.

For small geometric inductance $L$, such that $E_L \gg E_J$, the difference $\phi_-$ of the phases across the two junctions is confined in a very steep potential well and effectively fixed at the value of the external phase. Through this mechanism, the energy difference of the two logical states of the transmon $\epsilon = \sqrt{8 E_C E_J \cos(\phi_\mathrm{e}/2)}$ is tunable via the external flux.

By coupling the inductances of two qubits (labeled $\uparrow$ and $\downarrow$) to form a mutual inductance $M$ as shown in Fig.~\ref{fig:Qubit-Circuit}, we couple their potentials, and attain an effective interaction of the form $g^z \sigma^z_\uparrow \sigma^z_\downarrow$. The resulting coupling energy scale is
\begin{align}
g^z = - \frac{1}{16} \frac{M}{L} \tan (\frac{\phi_\uparrow}{2}) \tan ( \frac{\phi_\downarrow}{2}) \frac{\epsilon_\uparrow \epsilon_\downarrow}{E_L},
\end{align}
where $\phi_{\uparrow,\downarrow}$ are the phases associated with the external fluxes through the qubits. The coupling energy $g^z$ is now tunable through the external phases $\phi_{\uparrow,\downarrow}$, and even the sign of $g^z$ can be changed. Since $g^z \propto \epsilon_\uparrow \epsilon_\downarrow/E_L \propto E_J/E_L$, the coupling energy is very low. For it to be above the qubit linewidth one has to use transmon designs with not too low inductance. A possible experimental realization with substantial inductance is the concentric transmon qubit~\cite{Braum_concentric} (see Appx.~\ref{Appx:Experimental-Realization} for details).

To achieve the $XX$-type interaction between two transmons we suggest a coupling via the charge operators through the capacitances $C^x$ (see Fig.~\ref{fig:Qubit-Circuit}). For low coupling capacitances, $C^x \ll C$, we find an interaction of the form $g^x \sigma^x_j \sigma^x_{j+1}$ between nearest neighbor qubits with
\begin{align}
g^x = \frac{1}{4} \frac{C^x}{C} \, \epsilon \,.
\end{align}
Since $|g^x| \ll \epsilon$ the rotating wave approximation can be used, and we reproduce the exchange interaction in Eq.~\eqref{eq:Hamiltonian-Emulator}.%
\footnote{Note that this result is only valid in first order, and terms in $\mathcal O ((C^x/C)^2)$ have been dropped. They would lead to finite-range interactions decaying proportional to $(C^x/C)^k \sigma^x_j \sigma^x_{j+k}$. While it appears interesting to include interactions beyond nearest neighbors, one should note that such couplings in the qubit system do not map to a meaningful interaction in the fermionic system through the Jordan-Wigner transformation.}

\begin{figure}
\centering
\includegraphics[width=.66\columnwidth]{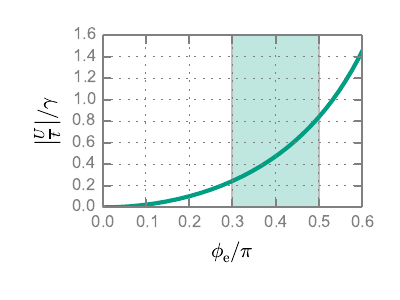}
\crop
\caption{
For a quantum simulator realized by transmon qubits,  the parameter $U/t$ of the Fermi-Hubbard model~\eqref{eq:Hamiltonian-Fermi-Hubbard-Model} to be emulated is plotted against the external phase $\phi_\mathrm{e}$ (assuming all external fluxes to be equal). The shaded area indicates the accessible range for $\phi_\mathrm{e}$, and $\gamma$ is a scaling factor (see main text).
}
\label{fig:U-t-Ratio}
\end{figure}

The AQS built in this way has an important feature: The tunability of $g^z$ through external fields enables to tune the ratio $U/t$ of the parameters of the Fermi-Hubbard model. Figure~\ref{fig:U-t-Ratio} displays a plot of $|U/t|$ over the external phase (which is assumed to be equal for all qubits). After absorbing the parameters of the circuit elements into the scaling factor $\gamma = \frac{C}{C^x} \frac{M}{L} \frac{\sqrt{8 E_C E_J}}{E_L}$ we find a universal behavior. The shaded area indicates (roughly) the allowed range of values of the external phase: If we choose $\phi_\mathrm{e}$ too small the value of $g^z$ will drop below the qubit linewidth, if we choose it too large, the transmons' susceptibility to noise increases. The sign of $U$ can be switched by reversing the external flux in either the upper or the lower chain of qubits, hence we can simulate repulsive or attractive on-site interactions.

Both types of couplings discussed here have energies that are small compared to the qubit energy splitting, i.e., $|g^x|,|g^z| \ll \epsilon$. For transmons this restriction is needed, since they are effectively anharmonic oscillators and the coupling strengths have to be weaker than the anharmonicity. Otherwise excitations of energy levels higher than the two logical states of the qubits play a role. This imposes a serious restriction on the parameters of the Fermi-Hubbard model~\eqref{eq:Fermi-Hubbard-Parameters}, namely it effectively amounts to $\mu \sim - \infty$. This means we can explore equilibrium properties of the system only in the limit of very low particle number.

We are, however, able to excite in a controlled way some of the qubits in the system and therefore initiate particles in the AQS. If the frequencies of the qubits to be excited are initially tuned far from the other ones in the AQS, and then tuned again to degeneracy adiabatically after the excitation, we should arrive in the ground state of the Fermi-Hubbard model with a given number of particles. Furthermore, after initiating specific excitations in the AQS, we can study the resulting non-equilibrium properties. This is especially interesting, as such properties are hard to treat either analytically or numerically. For this nonequilibrium dynamics there is no restriction on the filling of the system; i.e., we cover a wider range of parameters than in the equilibrium case.

\section{Initialization, Readout, and Control of the quality}

The fabrication process of the transmon Josephson junctions is not precise enough to guarantee the Josephson energies to be exactly equal (on the energy scale of the couplings). On the other hand, the qubit energies of the emulator's qubits have to be degenerate. For this reason, the proposed circuit in Fig.~\ref{fig:Qubit-Circuit} includes separate magnetic fields to tune individually each transmon. This extension should be manageable, adding one DC connection for each of the transmons to create the local fields. In addition, the individual tunability is useful for the initialization and readout, as well as for protocols to be used as tests of the quality of the simulation.

As an initialization scheme we suggest to couple a part of the qubit system capacitively to a transmission line as a feed line for external signals. By pairwise detuning the qubits in this part the $XX$ interaction between the transmons can be suppressed, which allows addressing every qubit individually through a resonant microwave pulse. Hence, one can excite specific qubits to initialize a desired configuration of excitations in the AQS. Note that the $ZZ$ interaction may only lead to an effective frequency shift for the excitation of a qubit, depending on whether the partner qubit is already excited or not. Bringing the transmons back to degeneracy turns the $XX$ exchange interaction back on, which will start the simulation.

Readout can be performed through the usual method of projective measurement in the dispersive regime. We couple resonators to the system, with frequencies far from the qubit frequencies during the simulation. Individual qubits are then tuned out of degeneracy to enter the dispersive regime of a resonator. This way, we can measure the time evolution of the operator $\sigma^z$ for each transmon.

An important issue to address is the quality of the results produced by the AQS~\cite{Hauck_Can_One_Trust_QS}. The AQS does not have error correction implemented. Deviations of the qubit system, where disorder effects cannot be avoided, from the ideal model Hamiltonian~\eqref{eq:Hamiltonian-Emulator} will modify conclusions to be drawn for the Fermi-Hubbard system, a problem which is largely unexplored. Here we propose tests which may help gaining confidence through several sorts of measurements. 

(i) One option is to check symmetries of the system. In our case, it is easy to see from the Hamiltonian~\eqref{eq:Hamiltonian-Fermi-Hubbard-Model} that the total number of excitations in the spin-up as well as the spin-down chain should be conserved separately. Furthermore, both chains should be equal, hence equal properties should be observed in both chains. In addition, for a sufficiently large system with negligible boundary effects, we  expect translational symmetry along the chains. 

(ii) If one creates fermions only in one of the chains (by exciting the corresponding qubits), one finds effectively a tight-binding model. In this case, the simulation results could be easily  compared to analytic results. 

(iii) Another option is to limit the system size of the AQS such that results could be compared to simulations on a classical computer. To do this, one does not need to build a small AQS. Rather, by tuning the coupling strengths it is possible to switch off the exchange interaction along the chains at a specific position. This will localize the excitations in one part of the simulator, and a comparison between the measurements of the smaller subsystem with numerical simulations could be possible.

Such measurements help in evaluating the quality of the emulator's results. Beyond such tests one should investigate the effects of disorder in the parameters of the emulator. First steps in this direction show for certain models (including the Fermi-Hubbard model) a remarkable stability of the AQS against disorder, which arises due to the symmetries ~\cite{Sarovar2016}. 

A limitation of the proposed emulator arises due to the limited coherence time of the qubits. But we estimate that qubit designs are available with sufficient quality to simulate the quantum state evolution on interesting, long time scales before decoherence dominates the dynamics. The effects of a bath coupled to a quantum emulator have been investigated in part~\cite{Schwenk2016} but need to be explored further. For a first estimate we have to compare the microscopic time scales with the decoherence time, or inversely, the coupling energies with the decoherence rate. For example, for the Xmon qubit coherence times of the order of $10\,\mathrm{\text\textmu s}$ to $100\,\mathrm{\text\textmu s}$ were reported~\cite{Martinis2013}, corresponding to rates from $100\,\mathrm{kHz}$ down to $10\,\mathrm{kHz}$. These rates are orders of magnitude smaller than the capacitive coupling energies ranging from $10\,\mathrm{MHz}$ to $100\,\mathrm{MHz}$. While this ratio looks very promising, we have to note that it is unclear how the required  magnetic coupling can be realized with the Xmon design. Therefore, we concentrated in this paper on the concentric transmon~\cite{Braum_concentric}, for which coherence times of the order of $10\,\mathrm{\text\textmu s}$ and capacitive coupling energies between $10\,\mathrm{MHz}$ and $100\,\mathrm{MHz}$ appear possible as well, and which allows for the magnetic coupling. First estimates for the magnetic coupling between two adjacent qubits yielded results only slightly above the decoherence rate, but  proper adjustments in the circuit should provide sufficiently strong coupling energies (see Appx.~\ref{Appx:Experimental-Realization}).

An actual quantity of interest in an emulation is a time- and space-dependent correlation function of the Fermi-Hubbard model on time scales which significantly exceed the microscopic time scales given by the inverse of the hopping matrix elements and the interaction strength. Accordingly, the correlation functions of the qubit circuit should be studied for times significantly exceeding the equivalent microscopic time scales determined by the inter-qubit couplings~\eqref{eq:Fermi-Hubbard-Parameters}. These time scales have to be compared with the coherence time of the $2n$-qubit double chain circuit. In the extreme case, if we are interested in long-range correlation functions, say of sites at opposite ends of the system, the relevant coherence time is reduced by a factor $2n$ as compared to the single-qubit coherence time. Thus, even with very promising ratios between single-qubit decoherence rates and coupling strengths, the study of long-range correlation functions is more difficult. But for not too large $n$ (say between 10 and 100) it should be possible to study with present technology the time dependence of correlation functions in an interesting regime. We finally note that with our proposed emulator we are in a much more favorable situation than with a digital quantum computer (without error  correction), where the simulation of the time evolution (according to the Trotter formula) requires executing a very large number of quantum gates making the limitations due to the finite coherence time much more serious.

\section{Conclusion}

A double chain of qubits with $XX$ and $ZZ$ couplings of neighboring qubits along and between the chains, respectively, can be mapped via the Jordan-Wigner transformation on a spin-full 1D Fermi-Hubbard model. The qubit system can thus be used to emulate the quantum properties of this model. It constitutes a different approach to, e.g., digital quantum emulation of the Fermi-Hubbard model with superconducting circuits~\cite{LasHeras2015,Barends2015}, or a recent effort with an emulator based on dopant atoms in a semiconductor~\cite{Salfi2015}. We analyzed different physical implementations of such analog quantum simulators. The conceptually simplest, with the broadest range of available parameters and highest flexibility is a realization based on an array of Josephson charge qubits. It would also allow simulations corresponding to half-filling of the fermionic problem. Unfortunately, Josephson  charge qubits suffer from the strong dependence on background charge fluctuations, which makes them difficult to handle in experiments. An alternative would be provided by using junctions based on phase slip processes~\cite{Mooij06,Mooij05,Astafiev}. For arrays built with such devices  disorder effects should be much weaker~\cite{Mooij15} which should make it worthwhile investing into this new technology. Because they are widely used nowadays, we discussed in detail tunable transmon qubits, where the $ZZ$ interaction arises due to an inductive coupling and the $XX$ interaction due to a capacitive interaction. This appears a promising approach as far as the experimental realization is concerned, and therefore it is discussed in more detail in Appx.~\ref{Appx:Experimental-Realization}. Although for this realization the parameter range is restricted, we could propose several interesting scenarios to be explored in an emulation, including protocols which can provide confidence in the results of the simulation through measurements of local operators.

\begin{acknowledgments}

We acknowledge fruitful discussions with A. Shnirman, A. Ustinov, A. Stehli, and M. {\v Z}nidari{\v c}. This work was supported by European Research Council (ERC) within consolidator Grant No. 648011, Deutsche Forschungsgemeinschaft (DFG) within Project No. \mbox{WE4359/7-1}. J. Brau\-m\"ul\-ler acknowledges financial support by the Lan\-des\-graduiertenf\"orderung (LGF) of the federal state Baden-W\"urttemberg.

\end{acknowledgments}

\section*{Appendices}

\begin{appendix}

\section{Tunable transmon}
\label{Appx:Tunable-Transmon}

\begin{figure}
\centering
\includegraphics[width=.66\columnwidth]{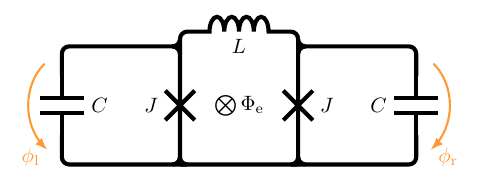}
\caption{
Circuit diagram of a tunable transmon with Josephson junctions J, the capacitances
C in parallel, and the intrinsic inductance L of the loop. The phase differences across the junctions
(and capacitances) are labeled $\phi_\mathrm{l}$ and $\phi_\mathrm{r}$, $\Phi_\mathrm{e}$ denotes an external flux through the loop.
}
\label{fig:Tunable-Transmon}
\end{figure}

In order to demonstrate the $ZZ$ and $XX$ couplings between two transmons in the circuit shown in Fig.~\ref{fig:Tunable-Transmon}, we expound the model of tunable transmon qubits. The transmon consists of a loop of two Josephson junctions $J$ with critical current $I_\mathrm{c}$. The loop has an intrinsic geometric inductance $L$, with an external magnetic flux $\Phi_\mathrm{e}$ passing through the loop. A large capacitance $C$ is shunted to each junction. The phases across the left and right junctions are denoted $\phi_\mathrm{l}$ and $\phi_\mathrm{r}$, and we introduce the external phase $\phi_\mathrm{e} = \frac{2e}{\hbar} \Phi_\mathrm{e}$. Using the addition theorem for the cosine potentials and defining $\phi = \frac{1}{2}( \phi_\mathrm{l} + \phi_\mathrm{r} )$, and $\phi_- = \frac{1}{2}( \phi_\mathrm{l} - \phi_\mathrm{r} )$ the Hamiltonian of the circuit reads
\begin{align} \label{eq:Transmon-Hamiltonian-Full}
H_\mathrm{T} &= 2E_C (N^2 + N_-^2) - 2 E_J \cos(\phi)\cos(\phi_-) \nonumber \\
&\quad + \frac{E_L}{2} (\phi_- - \frac{\phi_\mathrm{e}}{2})^2,
\end{align}
where we introduced the canonical charge number operators $N$, and $N_-$, as well as the energies $E_C = \frac{e^2}{2C}$, $E_J = \frac{\hbar}{2e} I_\mathrm{c}$, and $E_L = \frac{\hbar^2}{e^2 L}$.

The geometric inductance $L$ is very small, leading to $E_L \gg E_J$. Hence, we can neglect excitations of the degree of freedom associated with $\phi_-$. We fix $\phi_-$ at the minimum of its steep potential, resulting in the effective transmon Hamiltonian
\begin{align}
H_\mathrm{T,eff} = 2 E_C N^2 - 2 E_J \cos(\frac{\phi_\mathrm{e}}{2})\cos(\phi).
\end{align}

In the transmon limit, $E_J \gg E_C$, we can treat this as an anharmonic oscillator. For the lowest-lying states we can expand the cosine up to fourth order  and obtain a Duffing oscillator
\begin{align}
H_\mathrm{T,eff} = \epsilon (a^\dagger a + \frac{1}{2} ) - 2 E_J \cos( \frac{\phi_\e}{2} ) - \frac{E_C}{24} (a^\dagger - a)^4
\end{align}
with $\epsilon = \sqrt{8 E_C E_J \cos(\phi_\e /2)}$, and
\begin{align}
\phi &= - \mathrm{i} \frac{1}{\sqrt{2}} \Big( \frac{2E_C}{E_J \cos (\frac{\phi_\mathrm{e}}{2})} \Big)^\frac{1}{4} (a^\dagger -a), \\
N &= \frac{1}{\sqrt{2}} \Big( \frac{E_J \cos (\frac{\phi_\mathrm{e}}{2})}{2E_C} \Big)^\frac{1}{4} (a^\dagger + a).
\end{align}

A projection on the first two states casts the operators into the qubit basis, which yields (up to a constant contribution)
\begin{align}
H_\mathrm{T,eff} = \frac{1}{2} \epsilon \sigma^z,
\end{align}
with
\begin{align}
\phi &= \frac{1}{\sqrt{2}} \Big( \frac{2E_C}{E_J \cos (\frac{\phi_\mathrm{e}}{2})} \Big)^\frac{1}{4} \sigma^y, \\
N &= \frac{1}{\sqrt{2}} \Big( \frac{E_J \cos (\frac{\phi_\mathrm{e}}{2})}{2E_C} \Big)^\frac{1}{4} \sigma^x \label{eq:Transmon-N}, \\
\cos(\phi) &\approx 1 - \frac{\phi^2}{2} \nonumber \\
&= \Big(1- \frac{1}{2} \sqrt{\frac{2E_C}{E_J \cos (\frac{\phi_\mathrm{e}}{2})}} \Big) \mathbbm{1} -\frac{1}{4} \sqrt{\frac{2E_C}{E_J \cos (\frac{\phi_\mathrm{e}}{2})}} \sigma^z. \label{eq:Transmon-Cosine-Potential}
\end{align}

\section{\texorpdfstring{$ZZ$}{ZZ} coupling}
\label{Appx:ZZ-Coupling}

To achieve a  $ZZ$ coupling between two transmons, we make use of the previously ignored degree of freedom associated with $\phi_-$ in Eq.~\eqref{eq:Transmon-Hamiltonian-Full}. It couples to the cosine potential of the qubit states, thus to $\sigma^z$ (see Eq.~\eqref{eq:Transmon-Cosine-Potential}). By coupling the loop inductances of two transmons to form a mutual inductance $M$, the $\phi_-$'s of the transmons interact with each other, mediating a $ZZ$ interaction between the qubits.

For two inductances $L$ coupled to form a mutual inductance $M$, we can express $M=k_M L$, with $k_M \in (0,1)$ being a measure how close the inductances are coupled. We denote the currents through each inductance $I_{(1,2)}$, define the vector $\bm{I} = (I_1, I_2)$, and the matrix
\begin{align}
\bm{L} = \begin{pmatrix}
L & M \\
M & L
\end{pmatrix}.
\end{align}
For the magnetic fluxes through the inductances we write $\Phi_{(1,2)}$, and introduce $\bm{\Phi} = (\Phi_1, \Phi_2)$. With $\bm{\Phi} = \bm{L} \bm{I}$, the potential energy $V_M$ of the mutual inductance reads
\begin{align} \label{eq:Mutual-Inductance-Potential}
V_M &= \frac{1}{2} \bm{I}^\mathrm{T} \bm L \bm I = \frac{1}{2} \bm \Phi^\mathrm{T} \bm L^{-1} \bm \Phi \nonumber \\
&= \frac{1}{1-k_M^2} \frac{1}{2L} (\Phi_1 - \Phi_2)^2 + \frac{1}{1+k_M} \frac{1}{L} \Phi_1 \Phi_2.
\end{align}

\begin{figure}
\centering
\includegraphics[width=.66\columnwidth]{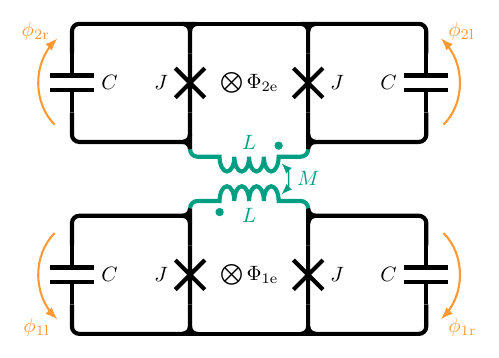}
\caption{
Circuit diagram of two inductively coupled transmons, where their inductances $L$ form a mutual inductance $M=k_M L$ with $k_M \in (0,1)$. The transmons should be built identically with Josephson junctions $J$ and shunt capacities $C$, but with individually controlled external fluxes $\Phi_{1\e}$ and $\Phi_{2\e}$ for tuning. $\phi_{(1,2)\mathrm l}$ and $\phi_{(1,2) \mathrm r}$ denote the phase differences across the junctions (and capacitances).
}
\label{fig:Inductively-Coupled-Transmons}
\end{figure}

Now we will derive the ZZ interaction for the transmon from Appx.~\ref{Appx:Tunable-Transmon}. Consider two transmons coupled by their inductances $L$ forming a mutual inductance $M=k_M L$ ($k_M \in (0,1)$), with the phases across the junctions (and capacitances) of the two transmons being denoted $\phi_{(1,2) \mathrm l}$ and $\phi_{(1,2) \mathrm r}$ as shown in Fig.~\ref{fig:Inductively-Coupled-Transmons}. For the external magnetic fluxes $\Phi_{(1,2)\mathrm{e}}$ we introduce the phases $\phi_{(1,2)\e} = \frac{2e}{\hbar} \Phi_{(1,2)\mathrm{e}}$. The Lagrangian of this setup reads (using Eq.~\eqref{eq:Mutual-Inductance-Potential}):
\begin{align}
\mathcal{L}^z &= \Big( \frac{\hbar}{2e} \Big)^2 \frac{1}{2} C ( \dot \phi_{1 \mathrm l}^2 + \dot \phi_{1 \mathrm r}^2 + \dot \phi_{2 \mathrm l}^2 + \dot \phi_{2 \mathrm r}^2) \nonumber \\
&\quad + \frac{\hbar}{2e} I_c ( \cos(\phi_{1 \mathrm l}) + \cos(\phi_{1 \mathrm r}) + \cos(\phi_{2 \mathrm l}) +\cos(\phi_{2 \mathrm r}) ) \nonumber \\
&\quad - \Big( \frac{\hbar}{2e} \Big)^2 \frac{1}{2L} \nonumber \\
&\quad\quad \Big( \frac{1}{1-k_M^2} \big( (\phi_{1 \mathrm l} - \phi_{1 \mathrm r} - \phi_{1\e}) - (\phi_{2 \mathrm l} - \phi_{2 \mathrm r} - \phi_{2\e}) \big)^2 \nonumber \\
&\quad\quad + \frac{2}{1+k_M} (\phi_{1 \mathrm l} - \phi_{1 \mathrm r} - \phi_{1\e}) (\phi_{2 \mathrm l} - \phi_{2 \mathrm r} - \phi_{2\e})\Big).
\end{align}
Defining $\phi_{(1,2)} := \frac{1}{2} (\phi_{(1,2) \mathrm l} + \phi_{(1,2) \mathrm r})$ as well as $\phi_\pm := \frac{1}{2} (( \frac{\phi_{1 \mathrm l} - \phi_{1 \mathrm r}}{2} - \frac{\phi_{1\e}}{2}) \pm ( \frac{\phi_{2 \mathrm l} - \phi_{2 \mathrm r}}{2} - \frac{\phi_{2\e}}{2}) )$, using that the external fields are constant, meaning $\dot \phi_{(1,2)\e} = 0$, and using the addition theorem $\cos ( \phi_{(1,2)\mathrm l} ) + \cos (\phi_{(1,2) \mathrm r}) = 2 \cos ( \frac{\phi_{(1,2) \mathrm l} + \phi_{(1,2) \mathrm r }}{2} ) \cos ( \frac{\phi_{(1,2) \mathrm l} - \phi_{(1,2) \mathrm r }}{2} )$ for the cosine terms one can find the Hamiltonian
\begin{align}
H^z &= 2 E_C ( N_1^2 + N_2^2 + \frac{1}{2} N_+^2 + \frac{1}{2} N_-^2 ) \nonumber \\
&\quad + \frac{1}{2} E_L \xi_+ \phi_+^2 + \frac{1}{2} E_L \xi_- \phi_-^2 \nonumber \\
&\quad - 2 E_J \cos(\phi_1) \cos(\frac{\phi_{1\e}}{2} + \phi_+ + \phi_-) \nonumber \\
&\quad - 2 E_J \cos(\phi_2) \cos(\frac{\phi_{2\e}}{2} + \phi_+ - \phi_-),
\end{align}
with the canonical charge numbers $N_{(1,2,\pm)}$, the energies $E_C$, $E_J$, and $E_L$ from Appx.~\ref{Appx:Tunable-Transmon} and the main text, as well as $\xi_+ = \frac{2}{1+k_M}$ and $\xi_-= \frac{4}{1-k_M^2} - \frac{2}{1+k_M}$.

Recalling that $L$ is very small, we find $E_L$ to be very large (compared to $E_J$). We can therefore ignore terms proportional to $E_J \phi_\pm^2$ towards those proportional to $E_L \phi_\pm^2$, which justifies expanding the cosines for small $\phi_\pm$ giving
\begin{align}
H^z &= 2 E_C ( N_1^2 + N_2^2 + \frac{1}{2} N_+^2 + \frac{1}{2} N_-^2 ) \nonumber \\
&\quad + \frac{1}{2} E_L \xi_+ \phi_+^2 + \frac{1}{2} E_L \xi_- \phi_-^2 \nonumber \\
&\quad - 2 E_J \cos(\phi_1) \big( \cos(\frac{\phi_{1\e}}{2}) - \sin(\frac{\phi_{1\e}}{2}) (\phi_+ +\phi_-) \big) \nonumber \\
&\quad - 2 E_J \cos(\phi_2) \big( \cos(\frac{\phi_{2\e}}{2}) - \sin(\frac{\phi_{2\e}}{2}) (\phi_+ - \phi_-) \big).
\end{align}

We can identify the qubit energy terms $2 E_C N_{(1,2)}^2 - 2 E_J \cos(\frac{\phi_{(1,2)\e}}{2}) \cos(\phi_{(1,2)}) = \frac{1}{2} \epsilon_{(1,2)} \sigma_{(1,2)}^z$, harmonic oscillators $E_C N_\pm^2 + \frac{1}{2} E_L \xi_\pm \phi_\pm^2 = \hbar \omega_\pm ( a_\pm^\dagger a_\pm + \frac{1}{2} )$, with $\omega_\pm = \frac{1}{\hbar} \sqrt{\xi_\pm 2 E_C E_L}$, and therefore $\phi_\pm = \frac{1}{\sqrt{2}} (\frac{2 E_C}{\xi_\pm E_L})^{\frac{1}{4}} (a_\pm^\dagger + a_\pm)$. We know that $\cos(\phi_{(1,2)})$ are diagonal in the qubit basis (see Eq.~\eqref{eq:Transmon-Cosine-Potential}), hence $\cos(\phi_{(1,2)}) = \alpha^z_{(1,2)} \sigma_{(1,2)}^z + \beta_{(1,2)} \mathbbm{1}$. Following the above argument that $\phi_\pm$ is very small, we neglect terms proportional to $E_J \mathbbm{1} \phi_\pm \approx 0$ and also leave out constant terms, resulting in
\begin{align} \label{eq:sz-Pre-Displacement}
H^z &= \frac{1}{2} \epsilon_1 \sigma_1^z + \frac{1}{2} \epsilon_2 \sigma_2^z + \hbar \omega_+ a_+^\dagger a_+ + \hbar \omega_- a_-^\dagger a_- \nonumber \\
&\quad + (g_{1+} \sigma_1^z + g_{2+} \sigma_2^z ) ( a_+^\dagger + a_+ ) \nonumber \\
&\quad + (g_{1-} \sigma_1^z - g_{2-} \sigma_2^z ) ( a_-^\dagger + a_- ),
\end{align}
with $g_{(1,2)\pm} = 2 E_J \alpha^z_{(1,2)} \sin(\frac{\phi_{(1,2)\e}}{2}) \frac{1}{\sqrt{2}} (\frac{2 E_C}{\xi_\pm E_L})^{\frac{1}{4}}$.

For the next step, we take the displacement operators $D_\pm(d) = \e^{ d a_\pm^\dagger - d^\dagger a_\pm}$, and define the unitary operator
\begin{align}
U = D_+(-\frac{g_{1+} \sigma_1^z + g_{2+} \sigma_2^z }{\hbar \omega_+} ) D_-(-\frac{g_{1-} \sigma_1^z - g_{2-} \sigma_2^z}{\hbar \omega_-} ).
\end{align}
By again neglecting constant terms, we transform the Hamiltonian to
\begin{align} \label{eq:U-Displacement}
U^\dagger H^z U &= \frac{1}{2} \epsilon_1 \sigma_1^z + \frac{1}{2} \epsilon_2 \sigma_2^z \nonumber \\
&\quad + \hbar \omega_+ a_+^\dagger a_+ + \hbar \omega_- a_-^\dagger a_- + g_z \sigma_1^z \sigma_2^z,
\end{align}
with $g^z = - 2 (\frac{ g_{1+} g_{2+}}{\hbar \omega_+} - \frac{ g_{1-} g_{2-}}{\hbar \omega_-})$. Because of the small inductance of the transmons, the energy $\hbar \omega_\pm$ is very big compared to the qubit energies such that the oscillators' excitation can be ignored again; they are just needed to mediate an effective interaction between the qubits.

Finally, combining the findings of this section, the convenient identity $(\frac{1}{\xi_+} - \frac{1}{\xi_-}) = k_M$, and the results from Appx.~\ref{Appx:Tunable-Transmon} one can check that for our effective Hamiltonian
\begin{align}
H^z_{\mathrm{eff}} = \frac{1}{2} \epsilon_1 \sigma^z_1 + \frac{1}{2} \epsilon_2 \sigma^z_2 + g_z \sigma^z_1 \sigma^z_2,
\end{align}
it holds that
\begin{align} \label{eq:gz-Mutual-Inductance}
g^z = -\frac{k_M}{16} \tan (\frac{\phi_{1\e}}{2} ) \tan (\frac{\phi_{2\e}}{2} ) \frac{\epsilon_1 \epsilon_2}{E_L}.
\end{align}

\begin{figure}
\centering
\includegraphics[width=.66\columnwidth]{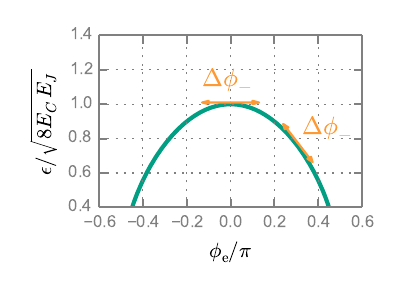}
\crop
\caption{
The transmon's level splitting $\epsilon$ depends on the phase $\phi_\mathrm{e}$ related to the external flux $\Phi_\mathrm{e} = \frac{\hbar}{2e} \phi_\mathrm{e}$. Fluctuations $\Delta \phi_-$ of the phase across the inductance of the transmon vary the energy splitting by an amount which depends on the slope of $\epsilon$.
}
\label{fig:Potential-Coupling}
\end{figure}

The coupling strength is tunable through the external phases $\phi_{(1,2)\e}$, including the possibility of a sign change. This results from the fact, that the interaction is mediated through displacements of the coupled phases $\phi_{(1,2)-}$. In a transmon, the extent to which a displacement of $\phi_-$ affects $\epsilon$ depends on the external flux (as illustrated  in Fig.~\ref{fig:Potential-Coupling}). Hence, the coupling strength is affected by the value of the external fluxes.

\section{\texorpdfstring{$XX$}{XX} coupling}
\label{Appx:XX-Coupling}

\begin{figure}
\centering
\includegraphics[width=\columnwidth]{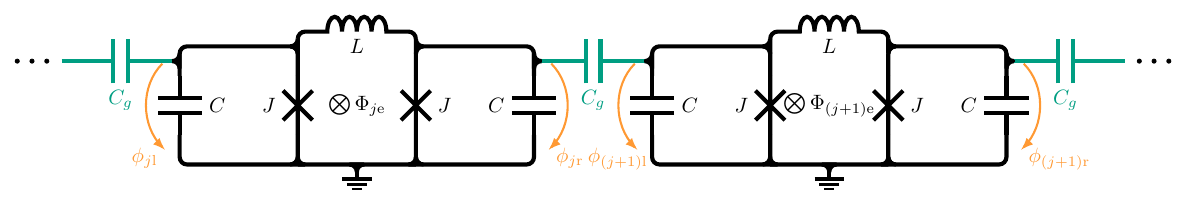}
\caption{
Circuit diagram of a chain of coupled identical transmons, with Josephson junctions $J$ (with critical current $I_\mathrm{c}$), shunt capacitances $C$, and loop inductance $L$. They are individually tunable though external fluxes $\Phi_{j\e}$, and the phase differences across the junctions (and capacitances) of the $j^\mathrm{th}$ qubit are denoted $\phi_{j \mathrm l}$ and $\phi_{j \mathrm r}$. The transmons are coupled by capacitances $C^x$ between neighboring transmons.
}
\label{fig:sx-Coupling-Capacitances}
\end{figure}

An $XX$ coupling arises since in a transmon the charge operator is proportional to $\sigma^x$ (see Eq.~\eqref{eq:Transmon-N}). We therefore suggest coupling the $n$ identical tunable transmons through capacitances $C^x$ as depicted in Fig.~\ref{fig:sx-Coupling-Capacitances}. The phase differences across the junctions (and capacitances) of the $j^\mathrm{th}$ qubit are denoted $\phi_{j \mathrm l}$ and $\phi_{j \mathrm r}$, and the external flux through the qubits are given by $\Phi_{j\e} = \frac{\hbar}{2e} \phi_{j\e}$.

Introducing $\phi_{j} = \frac{1}{2} ( \phi_{j \mathrm l} + \phi_{j \mathrm r})$, and $\phi_{j-} = \frac{1}{2} ( \phi_{j \mathrm l} - \phi_{j \mathrm r})$ we can rewrite the Lagrangian as
\begin{align}
\mathcal L^x  &= \sum_{j=1}^n \bigg( \Big( \frac{\hbar}{2e} \Big)^2 C (\dot{\phi}_{j}^2 + \dot{\phi}_{j-}^2) + \frac{\hbar}{e} I_c \cos(\phi_{j-}) \cos(\phi_{j}) \nonumber \\
&\quad \phantom{\sum} - \Big( \frac{\hbar}{2e} \Big)^2 \frac{2}{L} (\phi_{j-} - \frac{\phi_{j\e}}{2})^2 \bigg) \nonumber \\
&\quad + \sum_{j=1}^{n-1} \Big( \frac{\hbar}{2e} \Big)^2 \frac{1}{2} C^x \nonumber \\
&\quad \phantom{\sum} \big( (\dot{\phi}_{j} - \dot{\phi}_{j-}) - (\dot{\phi}_{(j+1)} + \dot \phi_{(j+1)-}) \big)^2,
\end{align}
and, regarding Appx.~\ref{Appx:Tunable-Transmon}, already fixing $\phi_{j-} = \phi_{j\e}/2$, $\dot \phi_{j-} = 0$ due to the steep potential created by the small value of $L$, we find
\begin{align}
\mathcal L^x  &= \sum_{j=1}^n \bigg( \Big( \frac{\hbar}{2e} \Big)^2 (C + C^x) \dot{\phi}_{j}^2 + \frac{\hbar}{e} I_c \cos(\frac{\phi_{j\e}}{2}) \cos(\phi_{j}) \bigg) \nonumber \\
&\quad - \sum_{j=1}^{n-1} \Big( \frac{\hbar}{2e} \Big)^2 C^x \dot{\phi}_{j} \dot{\phi}_{j+1} - \Big( \frac{\hbar}{2e} \Big)^2 \frac{1}{2} C^x (\dot \phi_{1}^2 + \dot \phi_{n}^2).
\end{align}

We denote $\tilde C = C + C^x$ and $\lambda = \frac{C^x}{2 \tilde C} \in (0, \frac{1}{2})$, also the tridiagonal $n \times n$ matrix
\begin{align}
\bm A_\lambda = \begin{pmatrix}
1 & \lambda & 0 & \cdots \\
\lambda & 1 & \lambda & \cdots \\
0 & \lambda & 1 & \cdots \\
\vdots & \vdots & \vdots & \ddots
\end{pmatrix},
\end{align}
with ones on the diagonal and $\lambda$ on the off-diagonals. Defining the vector $\bm \phi = (\phi_1, \dots, \phi_n)$ allows us to rewrite the Lagrangian as
\begin{align}
\mathcal L^x = \Big( \frac{\hbar}{2e} \Big)^2 \tilde C \dot{\bm\phi}^\mathrm{T} \bm A_\lambda \dot{\bm\phi}
+\sum_{j=1}^n \frac{\hbar}{e} I_c \cos(\frac{\phi_{j\e}}{2}) \cos(\phi_j),
\end{align}
where we neglected the boundary term $\frac{1}{2} C^x (\dot \phi_1^2 + \dot \phi_n^2)$. This is justified, as we will focus on the limit of small coupling, i.e., $C^x \ll C$. This limit is needed to apply the rotating wave approximation, as discussed in the main text.

Since $|\lambda| < \frac{1}{2}$, the matrix $\bm A_\lambda$ is invertible with $\bm A_\lambda^{-1} = \bm A_{-\lambda} + \mathcal O (\lambda^2)$. In the limit of weak coupling, we can neglect terms of order $\mathcal O (\lambda^2)$, which allows us to easily transform the Lagrangian to the Hamiltonian
\begin{align}
H^x &= 2 E_{\tilde{C}} \bm N^\mathrm{T} \bm A_{-\lambda} \bm N
- \sum_{j=1}^n 2 E_J \cos(\frac{\phi_{j\e}}{2}) \cos(\phi_j) \nonumber \\
&=  \sum_{j=1}^n \big( 2 E_{\tilde C} N_j^2 - 2 E_J \cos(\frac{\phi_{j\e}}{2}) \cos(\phi_j) \big) \nonumber \\
&\quad + \sum_{j=1}^{n-1} 4 E_{\tilde C} \lambda N_j N_{j+1}.
\end{align}

With the findings of Appx.~\ref{Appx:Tunable-Transmon}, we can identify the qubit energies $\tilde \epsilon_j = \sqrt{8 E_{\tilde C} E_J \cos(\phi_{j\e}/2)}$ and use Eq.~\eqref{eq:Transmon-N} to obtain $N_j = \alpha^x_j \sigma^x_j$ with $\alpha^x_j = \frac{1}{\sqrt{2}} (\frac{E_J \cos(\phi_{j\e}/2)}{2 E_{\tilde C}})^\frac{1}{4}$. This yields
\begin{align}
H^x =\sum_{j=1}^n \frac{1}{2} \tilde \epsilon_j \sigma^z_j + \sum_{j=1}^{n-1} g^x_j \sigma^x_j \sigma^x_{j+1},
\end{align}
where $ g^x_j = 2 E_{\tilde{C}} \lambda \alpha^x_j \alpha^x_{j+1} = \frac{1}{2} \lambda \sqrt{ \tilde \epsilon_j \tilde \epsilon_{j+1}}$.

For weak coupling we neglect the small change in the capacitative energy, especially since the transmons are tunable. Hence, we drop the tilde in the notation. Furthermore, we assume that the qubits are degenerate with respect to the energy $\epsilon$. The second order terms, $\mathcal O ((\frac{C^x}{C})^2)$, will also be dropped, yielding $\lambda = \frac{C^x}{2C}$. We thus obtain
\begin{align}
H^x =\sum_{j=1}^n \frac{1}{2} \epsilon \sigma^z_j + g^x \sum_{j=1}^{n-1} \sigma^x_j \sigma^x_{j+1},
\end{align}
with
\begin{align}
g^x = \frac{1}{4} \frac{C^x}{C} \epsilon.
\end{align}

\section{Experimental realization}
\label{Appx:Experimental-Realization}

\begin{figure}
\includegraphics[width=\columnwidth]{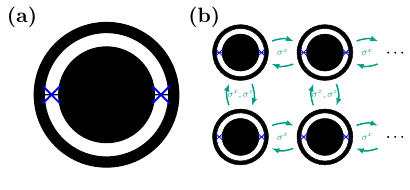}
\caption{
\textbf{(a)} Schematic of the concentric transmon qubit: It consists of a large central disk electrode ($230\,\mathrm{\text\textmu m}$ diameter) surrounded by a ring electrode, constituting the shunt capacitance for the two Josephson junctions interconnecting the electrodes. \textbf{(b)} Possible implementation of the simulator scheme depicted in Fig.~\ref{fig:Qubit-System} with concentric transmon qubits. In the vertical direction, the magnetic $ZZ$ coupling, mediated by a non-vanishing mutual inductance $M$, is relevant. The $XX$ coupling can effectively be suppressed by detuning vertically adjacent qubits. Transverse coupling is dominant in horizontal direction, where the $ZZ$ coupling is suppressed by exploiting the asymmetric gradiometry of the qubits.
}
\label{fig:conc}
\end{figure}

We explore the concentric transmon qubit \cite{Braum_concentric} as a potential candidate for the unit cell of the circuit diagram depicted in Fig.~\ref{fig:Qubit-Circuit}. It features a central disk island and a concentrically surrounding ring, constituting the shunt capacitance $C$ of the transmon and providing transversal $XX$ coupling to a neighboring device. The two qubit electrodes are interconnected by two Josephson junctions (see Fig.~\ref{fig:conc}(a)). The formed gradiometric SQUID allows for a fast tuning of the qubit frequency by a magnetic field gradient provided by an on-chip flux line. The geometric inductance of the large ring electrode provides a considerable magnetic dipole moment and thereby allows for an inductive $ZZ$ coupling to adjacent devices. By exploiting its rotational asymmetry with respect to the location of Josephson junctions, the concentric transmon architecture furthermore promises to satisfy the proposed simulator scheme since it allows for a site-selective engineering of the $ZZ$ coupling. The $XX$ coupling merely relies on the geometric dimensions of the qubit electrodes and is in first order isotropic. The effective dispersive coupling may be suppressed by a mutual frequency detuning. A proposal for the simulator scheme reproduced with concentric transmon qubits is schematically depicted in Fig.~\ref{fig:conc}(b).

Estimates indicate a small ratio $M/L$ between mutual and geometric inductance, resulting in a $ZZ$ coupling strength close to measured dephasing rates for the device. We are investigating design adaptations to increase the mutual inductance as well as the overall geometric inductance of the qubit circuit. One possible route to achieve this could be an interlocking, i.e., overlapping, of the ring electrodes of adjacent concentric transmons, leading to an increase in the mutual inductance while suppressing the capacitive coupling due to proximity to the central island. The concentric transmon may be considered as a viable starting point to further explore geometries with longitudinal $ZZ$ coupling eventually increased to an adequate strength.

\end{appendix}


%

\end{document}